\documentclass[aps,prl,twocolumn,groupedaddress]{revtex4-1}
\usepackage{graphicx}
\usepackage{amsmath}
\usepackage{color}

\usepackage{pdfpages}

\newcommand{\bra}[1]{\ensuremath{\left \langle {#1}\right|}}
\newcommand{\ket}[1]{\ensuremath{\left|{#1}\right \rangle}}
\newcommand{\braket}[2]{\ensuremath{\left \langle {#1} | {#2}\right \rangle}}
\newcommand{\abs}[1]{\lvert#1\rvert}

\begin{document}

\title{Demonstration of quantum entanglement between a single electron spin confined to an InAs quantum dot and a photon}

\author{J.\ R.\ Schaibley}
\author{A.\ P.\ Burgers}
\author{G.\ A.\ McCracken}
\author{L-M\  Duan}
\author{P.\ R.\ Berman}
\author{D.\ G.\ Steel}

\email{dst@umich.edu}
\affiliation{The H.\ M.\ Randall Laboratory of Physics, The University of Michigan, Ann Arbor, Michigan 48109-1040, USA}

\author{A.\ S.\ Bracker}
\author{D.\ Gammon}
\affiliation{The Naval Research Laboratory, Washington D.C. 20375, USA}

\author{L.\ J.\ Sham}
\affiliation{Department of Physics, The University of California, San Diego, La Jolla, California, 92093-0319, USA}

\date{\today}

\begin{abstract}
The electron spin state of a singly charged semiconductor quantum dot has been shown to form a suitable single qubit for quantum computing architectures with fast gate times. A key challenge in realizing a useful quantum dot quantum computing architecture lies in demonstrating the ability to scale the system to many qubits.  In this letter, we report an all optical experimental demonstration of quantum entanglement between a single electron spin confined to single charged semiconductor quantum dot and the polarization state of a photon spontaneously emitted from the quantum dot's excited state.  We obtain a lower bound on the fidelity of entanglement of $0.59\pm0.04$, which is 84\% of the maximum achievable given the timing resolution of available single photon detectors. In future applications, such as measurement based spin-spin entanglement which does not require sub-nanosecond timing resolution, we estimate that this system would enable near ideal performance. The inferred (usable) entanglement generation rate is $3 \times 10^{3}~\text{s}^{-1}$. This spin-photon entanglement is the first step to a scalable quantum dot quantum computing architecture relying on photon (flying) qubits to mediate entanglement between distant nodes of a quantum dot network.
\end{abstract}

\maketitle

A single electron spin confined to a charged semiconductor quantum dot (QD) can effectively serve as a single quantum storage device with fast information processing for quantum computing architectures \cite{loss_div_1998,imam_qubit_1999, sham_Cavity_photon}. QD architectures are excellent candidates for scalable quantum information applications since they are compatible with existing semiconductor processing infrastructure. In addition, site-controlled QD growth has been demonstrated \cite{ishikawa_site_1998, atkinson_site-controlled_2008}, and single QDs have been integrated with photonic crystal cavities \cite{badolato_deterministic_2005, gallo_integration_2008}, offering significant advantages of optically driven QD spins over other modern quantum information systems. In order to construct a scalable architecture, quantum information must be coherently transferrable between electron spin qubits in separate nodes.  The photons emitted from an excited, negatively charged QD (called a trion: a multi-particle state comprised of two electrons and one hole) provide an attractive messenger to carry this information.  Recently, optical initialization, rotation and readout of a single electron spin qubit in a single QD were accomplished, demonstrating the QD spin's usefulness as a single qubit \cite{atature_quantum-dot_2006,xu_fast_2007,press_complete_2008,kimspin2010}. Scaling the architecture to arbitrary size requires the ability to entangle the spin qubits of spatially distinct QDs, recently demonstrated by using the tunneling interaction between spatially adjacent QDs \cite{kim_ultrafast_2011}. One scaling approach that does not require local interactions instead uses photon qubits to entangle the QDs \cite{cabrillo_creation_1999, duan_long-distance_2001, sham2005photonQD, duan_colloquium:_2010}.  If the photons emitted from two QDs are indistinguishable, coincidence measurements can be performed on the emitted photons to probabilistically entangle the source QDs \cite{cabrillo_creation_1999, duan_long-distance_2001, moehring_quantum_2007,moehring_entanglement_2007}. The first step in protocols of this nature is establishing the entanglement between a single emitted photon and a single QD spin.

In this letter, we report entanglement between a single electron spin state confined to a single semiconductor QD and the polarization state of a photon that has been emitted spontaneously from the QD's excited state \cite{note1}.  The entanglement is verified by performing projective measurements on the entangled photon's polarization state and time correlating this detection with the resulting electron spin state of the QD in two bases.  The protocol follows established techniques in quantum information systems using single atoms and nitrogen vacancy centers in diamond \cite{blinov_observation_2004, volz_entanglement_2006, wilk_single-atom_2007,togan_quantum_2010}. This demonstration of entanglement represents a hybrid entanglement between an engineered quantum state and a traveling qubit and is integral to future applications using QDs in quantum information and scalable quantum computing applications. The validity of the approach used here and in other recent experiments \cite{wilk_single-atom_2007,togan_quantum_2010} has recently been justified theoretically. \cite{schaibley_effect_2012}.

The energy level structure of a single charged QD in the presence of an externally applied magnetic field (Voigt geometry) is shown in Fig. 1(a) with the corresponding optical selection rules \cite{xu_fast_2007}. In the experiment, the QD is initialized to a pure state via optical pumping, then excited to the $\ket{T_{x}-}$ trion state with a laser pulse, where it then decays to the two ground states with equal probability \cite{xu_fast_2007}. When the $\ket{T_{x}-}$ state decays, the horizontal (vertical) ($H$,$V$) polarization state of the emitted photon, collected along the $z$ axis, is correlated with the final state ($\ket{x+}$, $\ket{x-}$) of the QD.  Here, the electron ground state frequency splitting ($\Delta_{e}= 2 \pi \times 7.35~\text{GHz}$) is larger than the spontaneous emission rate ($10^{9} ~\text{s}^{-1} $), so a fast detector with timing resolution ($\tau_{r}$) of $48~\text{ps}$ FWHM is used to destroy the which-path information from the frequency mismatch of the two decay channels \cite{scully_quantum_1982, economou_unified_2005, volz_entanglement_2006, wilk_single-atom_2007, togan_quantum_2010, schaibley_effect_2012}.  The resulting state vector ($\ket{\Psi}$) of the system is,

\begin{align}
\ket{\Psi} =\frac{\ket{H}\ket{x+} -i\ket{V} \ket{x-}}{\sqrt{2}},
\end{align}
clearly reflecting the entanglement \cite{ economou_unified_2005}.

The state of the photon is measured with a single photon avalanche photodiode (SPAPD) after polarization analysis.  The measurement of the photon's polarization is correlated uniquely with a particular final state in the QD. A narrow bandwidth laser pulse reads out the resulting electron spin state by selectively scattering from only one of the ground states, mapping the QD spin state into a readout photon which is detected by another SPAPD.  The photon and spin measurements are analyzed based on their time correlated nature to reconstruct the state of the spin-photon system. First, we confirm that the detection of a $H$ ($V$) polarized photon is correlated with the $\ket{x+}$($\ket{x-}$) state of the QD. We then verify that the state is entangled by rotating both measurement bases by $\pi/2$ about the $y$ axis and showing that the measured state of the spin in the $z$ basis ($ \ket{z\mp} = \frac{ \ket{x+} \pm \ket{x-}}{\sqrt{2}}$) remains correlated with the detection of a circularly polarized photon ($\ket{\sigma\pm }=\frac{ \ket{H} \pm i \ket{V}}{\sqrt{2}} $). This is possible due to long coherence time of the QD spin state \cite{petta_coherent_2005,greilich_mode_2006,press_ultrafast_2010}.

\begin{figure}
\begin{center}
  \includegraphics[width=3.5in]{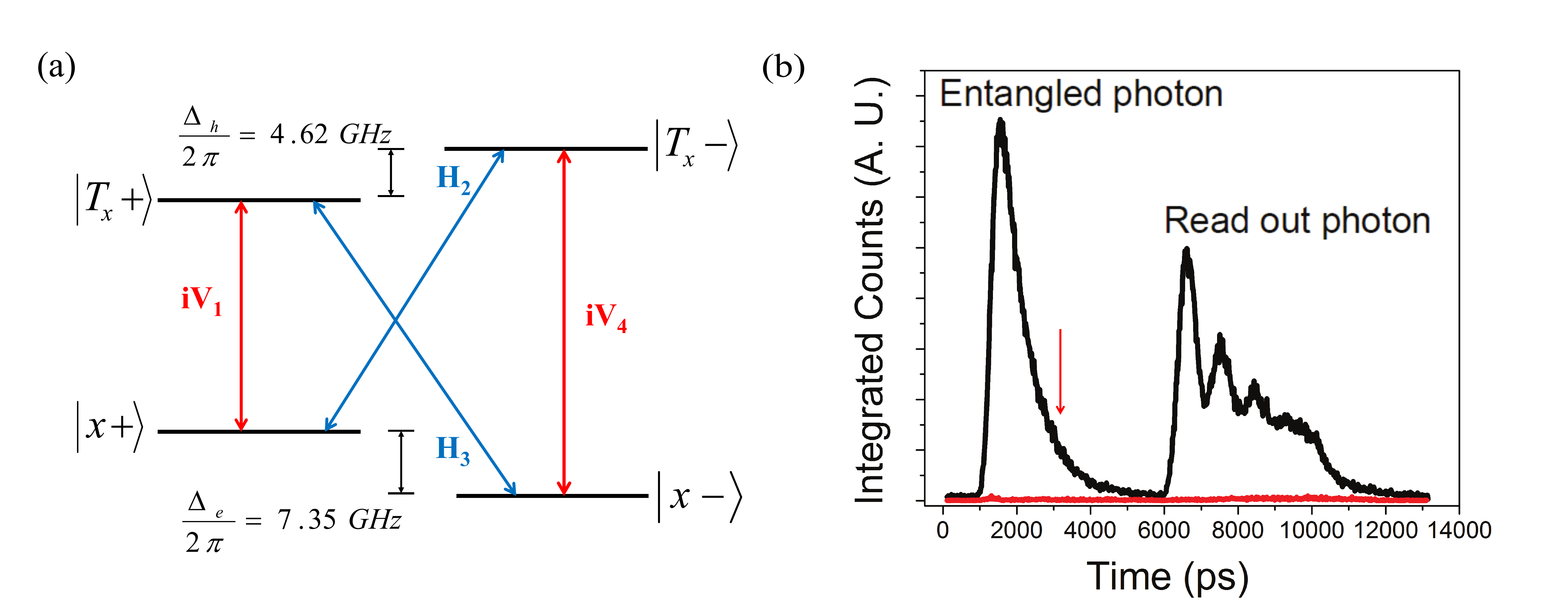}\\
  \caption{(a)~The effective four level system generated when a magnetic field is applied perpendicular to the QD growth axis. The selection rules are shown to be horizontal $H$ (vertical ($V$)) where the $i$ is included to illustrate the relative phase between the matrix elements.  The subscripts label the transitions in order of increasing energy.  The excited state (heavy hole) splitting ($\Delta_h/2\pi$), and the ground state (electron) splitting ($\Delta_e/2\pi$) are shown. (b)~Time histogram of integrated fluorescence showing QD emission (black) from the excitation and readout/intialization pulses. The red shows the background level when the QD bias is tuned off resonance with the excitation lasers.  The arrow indicates the temporal location of the rotation pulse used in the $z$ basis measurements.  }
\end{center}
\end{figure}

The system investigated is a single negatively charged InAs QD embedded in a GaAs Schottky diode heterostructure grown via molecular beam epitaxy. The characterization of QDs is discussed in detail in earlier work \cite{tischler_fine_2002,xu_fast_2007}.  Optical studies are performed at $\approx7~\text{K}$ with a combination of pulses from CW lasers produced by $\text{LiNbO}_{3}$ electro-optic modulators which are synchronized with a $76$ MHz mode-locked Ti:Sapphire laser. A $4~\text{ns}$ resonant laser pulse initializes to either the $\ket{x+}$ or $\ket{x-}$ state of the QD, and a resonant $250~\text{ps}$ ($\Theta_{trion}=\pi$ area) pulse selectively excites this state to $\ket{T_{x}-}$. The resulting spin state following spontaneous emission is then measured by a resonant state selective readout pulse (either $4~\text{ns}$ or $250~\text{ps}$).  For the rotated ($\ket{z\mp}$,$\ket{\sigma\pm}$) basis measurements, a $\approx2~\text{ps}$ ($\Theta_{spin}=\pi/2$ area) Raman pulse, red detuned by approximately $1~\text{meV}$, is used to rotate the $z$ basis state into an $x$ basis state prior to readout by the $4~\text{ns}$ measurement pulse\cite{press_complete_2008,kimspin2010}. The pulse widths and magnetic field are chosen to simultaneously allow for frequency selective state excitation, while at the same time keeping the ground state splitting small compared to the bandwidth of our detector. The entangled and readout photons are projected by a polarization analyzer and quarter-wave plate which is used either to convert back to linear polarization or to correct for birefringence in the cryostat's windows.  The QD emission is then coupled into a single mode fiber, split with a 50-50 fiber splitter and sent to two SPAPDs in a HBT-type setup \cite{brown_correlation_1956}. The photon arrival times are time tagged relative to the excitation pulses using a picosecond event timer. For the $z$ basis measurement, a fast timing SPAPD is used to measure the entangled photon's arrival time (timing jitter $48~\text{ps}$ FWHM) that sets the maximum observable spin precession rate (Zeeman splitting).  For this QD, that splitting corresponds to a magnetic field of $1.1~\text{T}$. For each photon projection axis ($H$,$V$,$\sigma+$,$\sigma-$), the excitation and rotation lasers were polarized orthogonally to the measurement axis. The QD emission is separated from the excitation lasers by a combination of polarization and spatial filtering. For the rotated ($\ket{z\mp}$, $\sigma\pm $) basis measurements, an air spaced etalon is used to further attenuate the detuned rotation pulse by $30~\text{dB}$. The rejection ratio of the narrow bandwidth pulses exceeds $70~\text{dB}$. The probability of false correlations contributing to our signal due to resonant excitation leak through is less than 0.02 for the $x$ basis measurements and less than 0.05 for the $z$ basis measurements. Due to the time correlated nature of the measurements, false correlations from detector dark counts are negligible. The setup's single channel detection efficiency ($DE$) is $ \approx 4 \times 10^{-5}$;  the detection efficiency of the fast timing resolution SPAPD required for the $z$ basis measurement is $ \approx 4 \times 10^{-6}$.

The experimental pulse sequences are shown in Fig. 2.  Six independent measurements are performed to obtain the conditional probabilities shown in Fig. 3. For the $H$ and $V$ measurements, four separate measurements are performed, one for each of the $x$ basis conditional probabilities (Fig. 3(a)).  For the $\sigma \pm$ measurements, two separate measurements are performed, each of which simultaneously measures two $z$ basis conditional probabilities (Fig. 3(b)).  In the first measurement, the correlation between a $H$ emitted photon and the $\ket{x+}$ state is established using a two pulse sequence where both pulses are linearly polarized with the vertical (horizontal) transitions (Fig. 2(a)).  The QD is initialized to $\ket{x-}$ with a $4~\text{ns}$ ($\Omega_{CW}/2\pi \approx 1~\text{GHz}$, where $\Omega_{CW}$ is the Rabi frequency) pulse tuned to the $V_{1}$ transition. Then a $250~\text{ps}$ pulse ($\pi$ area), tuned to the $V_{4}$ transition, excites the system to $\ket{T_{x}-}$, followed by spontaneous emission. We then correlate the final state of the QD with the polarization of the emitted photon. The next $4~\text{ns}$ initialization pulse also serves as a readout pulse for the state of the QD. It scatters a photon only if the QD is in the $\ket{x+}$ state. In the event that no photon is collected after the $250~\text{ps}$ pulse, the probability of detecting a readout photon is half as likely, since we have no information on the final state of the QD. In the second measurement, we then perform a negative correlation measurement between $H$ and $\ket{x-}$ by inserting an additional $250~\text{ps}$ (probe) pulse between the existing $250~\text{ps}$ (excitation) pulse and $4~\text{ns}$ pulse (which now serves only to re-initialize).  Here, upon detection of a $H$ polarized photon following the first $250~\text{ps}$ pulse, the spin is projected to $\ket{x+}$, so the second $250~\text{ps}$ probe pulse should not scatter any photons off the $\ket{x-}$ state (Fig. 2(b)). This pair of experiments is then repeated with initialization to $\ket{x+}$ using a $4~\text{ns}$ pulse tuned to the $H_{3}$ transition and a $250 ~\text{ps}$ pulse tuned to the $H_{2}$ transition.  In analogy with the first two measurements, we then establish the correlation between a $V$ emitted photon and $\ket{x-}$ or a negative correlation with $\ket{x+}$. We normalize the conditional probabilities by comparing the number of correlations between the entangled photons and those from the $4~\text{ns}$ or $250~\text{ps}$ readout pulse to the number of correlations between an entangled photon with a readout photon from temporally distant runs of the experiment (which corresponds to a probability of $0.5$ for a $\pi$ excitation pulse).
\begin{figure}
\begin{center}
  \includegraphics[width=3.5in]{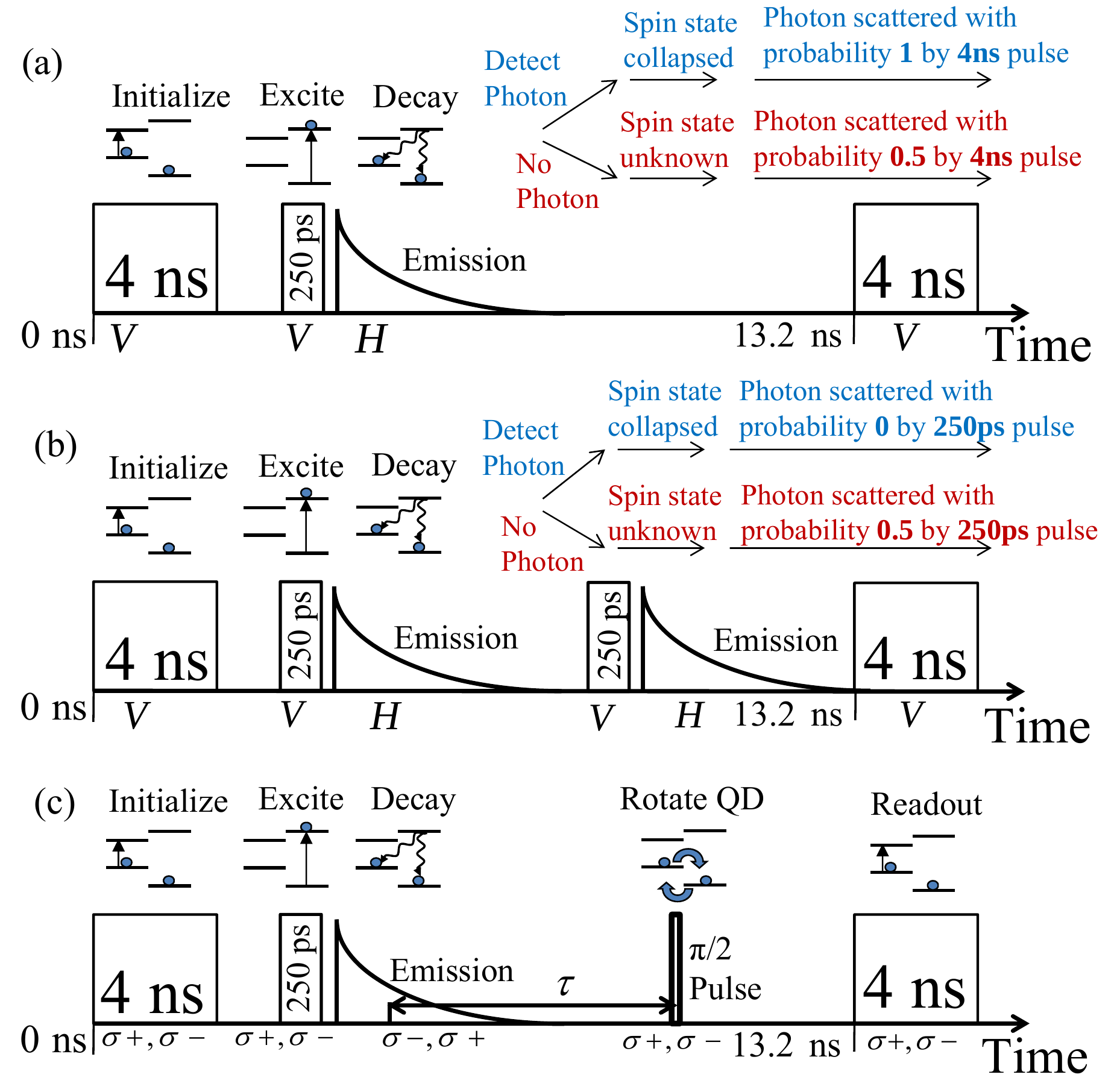}\\
  \caption{(a)~Pulse sequence used for $P(x+|H)$ measurement.  After initialization, a $250$ ps $\pi$ pulse excites to $\ket{T_x-}$.  Upon detection of a $H$ polarized photon, the spin state ideally collapses to $\ket{x+}$ where the population is read out by the next $4$ ns pulse. (b)~To show anti-correlation ($P(x-|H)$) in an independent measurement, a second $250$ ps $\pi$ pulse is used to readout the remaining $\ket{x-}$ population after a $H$ photon is detected. (c)~To verify the entanglement, we perform the correlation measurement in the rotated ($z$) basis.  Here, we excite with $\sigma\pm$ and detect $\sigma\mp$.  A detuned $\pi/2$ Raman  pulse is used after the $250$ ps pulse to rotate the spin coherence into a probability amplitude that is read out by the following $4$ ns pulse.  The photon detection time is binned relative to the Raman pulse ($\tau$) to observe the coincidence oscillations at the electron difference frequency.}
\end{center}
\end{figure}

An example of the time integrated emission from a positive correlation measurement is shown in Fig. 1(b). We measure the probability of recording coincident photons on each of the two SPAPDs during the same pulse and use this to correct the raw data. The corrected data are normalized requiring the sum of each pair to equal one \cite{corrs}. The corrected conditional probabilities calculated, shown in Fig. 3(a), are:  $P(x-|V)= 0.84 \pm 0.04$, $P(x+|V) = 0.16 \pm 0.01$, $P(x+|H)= 0.94 \pm 0.05$, and $P(x-|H)= 0.06 \pm 0.01$. The uncorrected values are: $P(x-|V)= 0.68 \pm 0.02$, $P(x+|V) = 0.25 \pm 0.02$, $P(x+|H)= 0.91 \pm 0.03$, and $P(x-|H)= 0.12 \pm 0.04$. We note that the primary source of error is off-resonant coupling of the laser pulses to the other trion state. This coupling is more pronounced in the $V$ configuration, where the lasers are driving the $H$ transitions which are the closest in energy, and is manifested in the lower fidelity of the $V$ measurement as well as the sum of the uncorrected conditional probabilities differing further from one.  This error is partially corrected by the subtraction method used to obtain the corrected values, but remains detrimental to the fidelity due to imperfect state initialization \cite{corrs}. The unintended excitation can in principle be removed by pulse shaping \cite{piermarocchi_theory_2002}.

\begin{figure}
\begin{center}
  \includegraphics[width=3.5in]{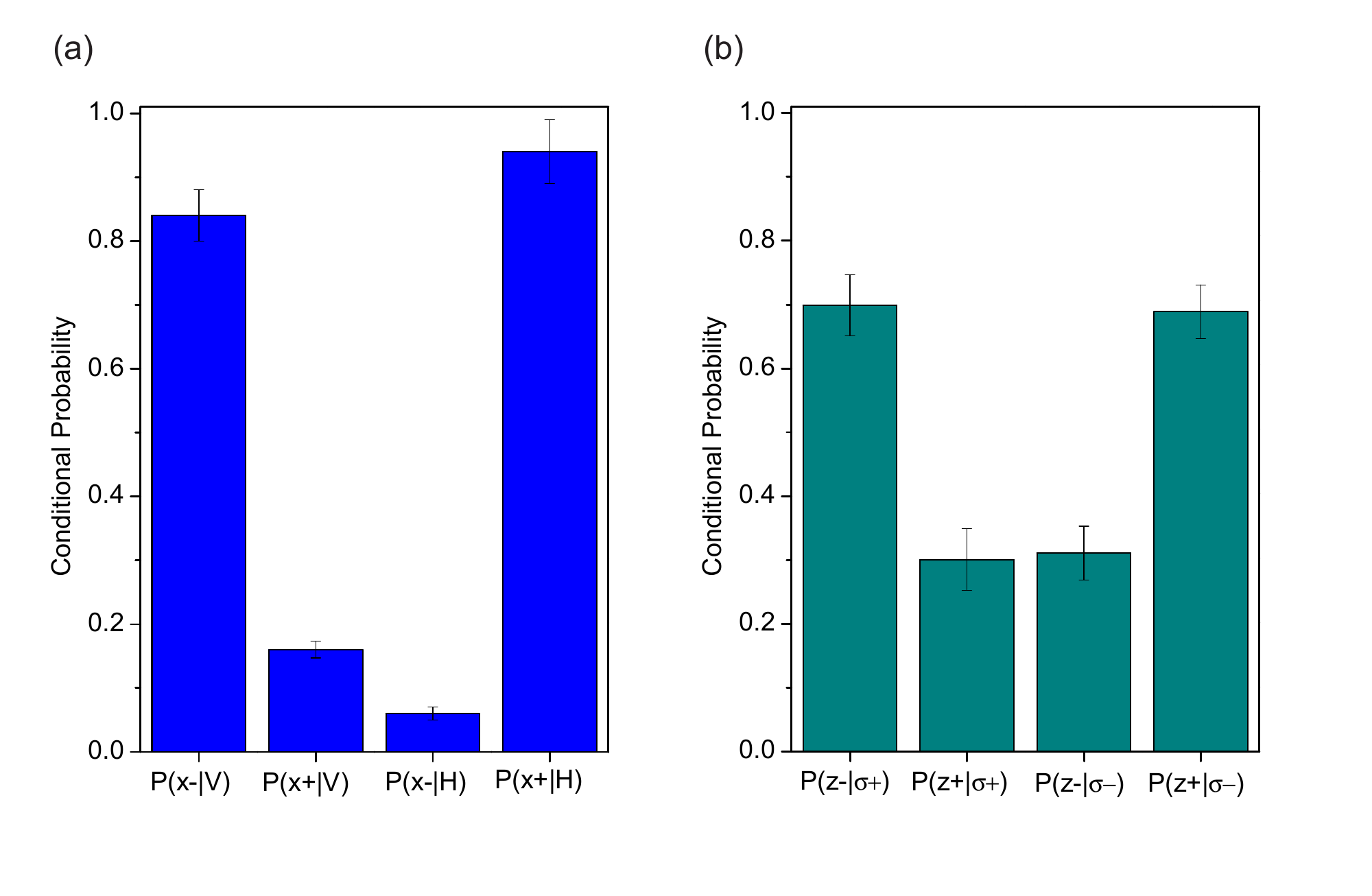}\\
  \caption{Conditional probabilities showing the correlated nature of the entangled spin-photon state in two bases. (a)~For the $H$,$V$ measurements, corrected data are shown. (b)~ For the $\sigma\pm$ measurements, the conditional probabilities are extracted from fits shown in Fig. 4.}
\end{center}
\end{figure}

In our final measurements, carried out using the $z$ basis (i.e., the rotated basis), the correlation is time dependent allowing for simultaneous measurement of two conditional probabilities. The $1.1~\text{T}$ magnetic field keeps the ground state precession period longer than the timing resolution of the fast timing resolution SPAPD while splitting the excited states sufficiently to allow frequency selective excitation since the circular polarized laser pulses can couple to either transition.  This will lower the fidelity of entanglement because of the reduced quality of initialization into a pure state.  For both measurements, the QD is initialized to $\ket{x-}$ with a 4 ns pulse tuned to the $V_{1}$ transition and then excited to the $\ket{T_{x}-}$ state with a $250~\text{ps}$ pulse resonant on the $V_{4}$ transition.  The excited QD decays to both lower spin states. The photon state is measured along $\sigma\pm$, which projects the QD spin to a superposition of $x$ basis states.  The spin state evolves according to Schr\"{o}dinger's equation until a time ($\tau$) later when a $\pi/2$ spin rotation pulse maps the coherence into an $x$ basis probability amplitude.  This is read out by a scattered photon during the next $4~\text{ns}$ pulse. The form of the signal, after dividing out by an exponential decay envelope, is
\begin{align}
\abs{\braket{x+} { {R_{\sigma\mp}}(\pi/2)U(\tau)\braket{\sigma\pm}{ \Psi}}}^{2}= \frac{1}{4} (1+\sin{\Delta_{e} \tau}),
\end{align}
where $R_{\sigma\mp}= \frac{1}{\sqrt{2}} (\ket{x+}\bra{x+} \pm i \ket{x+}\bra{x-} \pm i \ket{x-}\bra{x+} +  \ket{x-}\bra{x-})$, $U(\tau)$ is the time evolution operator, and $\Delta_{e}$ is the electron spin difference frequency.

Since, the radiative lifetime of the trion state ($\approx1~\text{ns}$) is longer than the spin precession period, the time $\tau$ varies randomly with an exponentially decaying probability.  Upon measurement of the entangled photon, the spin state is re-initialized to $\ket{z\pm}$, serving as a measure of the phase of the generated spin coherence. One can view the timing resolution requirement ($\tau_{r} < 2\pi/\Delta_{e}$) as a quantum-eraser effect, where the photon detection must be sufficiently achromatic to avoid measuring the frequency mismatch between the two decay paths \cite{scully_quantum_1982, economou_unified_2005, volz_entanglement_2006, wilk_single-atom_2007, togan_quantum_2010, schaibley_effect_2012}.  The data are shown in Fig. 4 along with fits of the first three periods to Eq. (2) using the experimentally determined spin difference frequency ($\Delta_{e}/2\pi =7.35~\text{GHz}$). From the fringe contrasts, we extract the conditional probabilities: $P(z-|\sigma+)=0.70 \pm0.05$, $P(z+|\sigma+)=0.30 \pm0.05$, $P(z-|\sigma-)=0.31 \pm0.04$, $P(z+|\sigma-)=0.69 \pm0.04$ .

\begin{figure}
\begin{center}
  \includegraphics[width=3.5in]{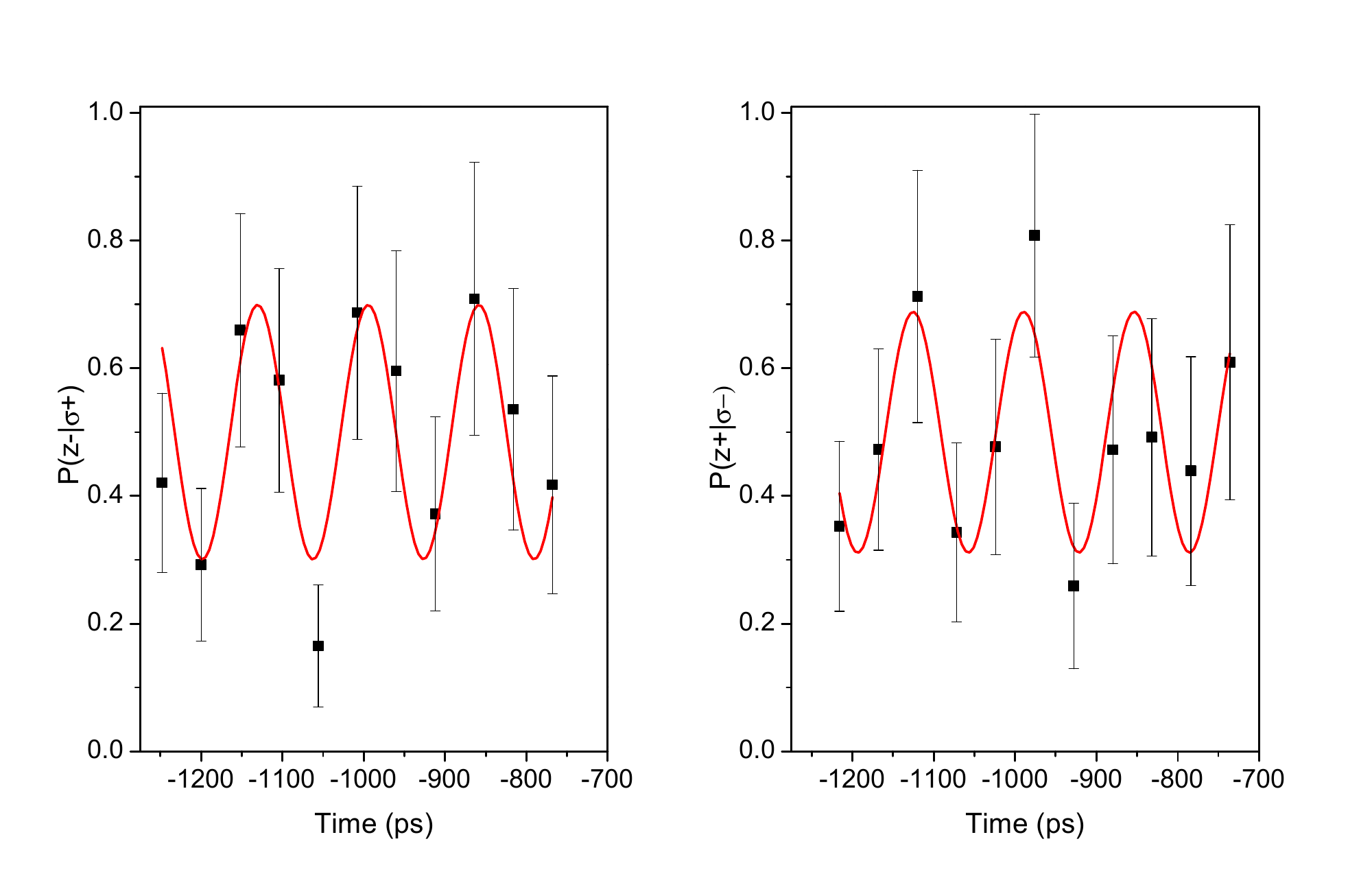}\\
  \caption{ Time resolved coincidence oscillations showing the QD spin coherence generated by projecting of the photon state onto $\sigma\pm$ for the left and right figures respectively. The time axis is taken relative to the QD spin rotation pulse which occurs at $t=0$. The first three periods of the normalized data are fit to Eq. (2) using the experimentally determined difference frequency ($7.35~\text{GHz}$). The data show fringe contrasts of $0.40 \pm 0.10 $ for $\sigma+$ and $0.38 \pm 0.08$ for $\sigma-$. Note that because we remove the exponential envelope by division, the relative noise increases with time.}
\end{center}
\end{figure}

We calculate a lower bound on the entanglement fidelity of $\mathcal{F}\geq 0.59 \pm 0.04$ using the expression $\mathcal{F}\geq 1/2 (\rho_{H x+,Hx+} + \rho_{V x-,V x-}-2\sqrt{\rho_{H x-,H x-}\rho_{V x+,V x+}} +\rho_{\sigma+ z-,\sigma+ z-}-\rho_{\sigma+ z+,\sigma+ z+} +\rho_{\sigma- z+,\sigma- z+}-\rho_{\sigma- z-,\sigma- z-})$ \cite{blinov_observation_2004}. Here, we note that $ 2\pi/\Delta_{e} \approx  2.8 \times \tau_{r}$, so the reduction in fringe contrast is limited almost entirely by instrumental convolution.  By convolving the theoretical signal with the detection system's instrument response function, and assuming a perfect correlation in the $x$ basis, we estimate our experimentally realizable fidelity to be $\approx0.7$, putting the measured fidelity bound at 84\% of the detector limited bound. The deviation from 100\% of the maximum achievable fidelity is primarily from imperfect state initialization which is most pronounced in the $V$ polarized ($x$ basis) measurements. \cite{corrs}.

For quantum information applications such as QD spin-spin entanglement mediated by spin-photon entanglement, QD spin-photon entanglement is essential \cite{moehring_entanglement_2007}. An important distinction of such a scheme is that the detector's timing resolution no longer plays a limiting role, allowing for higher magnetic fields, and therefore achievable fidelities approaching unity. The success rate of the $x$ ($z$) basis measurement is approximately $0.06 ~\text{s}^{-1}$ ($0.002 ~\text{s}^{-1}$); however, the entanglement generation rate is given by the rate of entangled photons detected which is $DE \times 76~\text{MHz} = 3 \times 10^{3} ~\text{s}^{-1}$. In a protocol similar to Moehring et al. \cite{moehring_entanglement_2007}, this would result in a spin-spin entanglement generation rate of approximately once per minute.  Efficient spin readout should be possible by using a QD molecule sample capable of non-destructive spin measurement \cite{kim_optical_2008}. A feasibility analysis of using intermediate spin-photon entanglement to mediate distant QD spin-spin entanglement is given in the supplemental material \cite{spin-spin-sup}.  Integrating these techniques has the potential to form a scalable QD spin architecture suitable for many quantum information applications.

After the submission of this work, two papers appeared in which results of a similar nature were reported  \cite{greve_quantum-dot_2012, gao_observation_2012}. A discussion comparing the physics of these measurements to our result is given in the supplemental material \cite{spin-spin-sup}.

Acknowledgements: This research is supported by NSF PHY0804114, PHY1104446; AFOSR FA9550-09-1-0457; DARPA FA9550-10-1-0534, FA8750-12-2-0333; ARO W911NF-08-1-0487 and MURI W911NF-09-1-0406.

\bibliography{spinphotonbib}

\begin{thebibliography}{37}
\bibitem [{not()}]{note1}%
  \BibitemOpen
  \href@noop {} {}\bibinfo {note} {Since submission of this paper, two papers
  have been published reporting entanglement between a QD spin and the photon
  polarization and a QD spin and the photon wavelength [36, 37]}\BibitemShut
  {NoStop}%
\end{thebibliography}



\section*{Supplementary material}

# Supplementary Information for "Demonstration of quantum entanglement between a single electron spin confined to an InAs quantum dot and a photon"

J. R. Schaibley, A. P. Burgers, G. A. McCracken, L-M Duan, P. R. Berman, and D. G. Steel*
*The H. M. Randall Laboratory of Physics, The University of Michigan, Ann Arbor, Michigan 48109-1040, USA*

A. S. Bracker and D. Gammon
*The Naval Research Laboratory, Washington D.C. 20375, USA*

L. J. Sham
*Department of Physics, The University of California,
San Diego, La Jolla, California, 92093-0319, USA*
(Dated: March 13, 2013)

## SPIN-SPIN ENTANGLEMENT

A primary application of quantum dot (QD) spin-photon entanglement is to mediate spin-spin entanglement between distant QDs by using post-selection [1–3]. Although there are many schemes which can use spin-photon entanglement to mediate spin-spin entanglement [1–6], we focus on the feasibility of implementing a type-II protocol similar to Moehring et al. [3], since many of the challenges are shared between the protocols. In such protocols, two spin-photon entangled states are generated, and the photons from each of the atoms are interfered in a Hong-Ou-Mandel (H.O.M.) interferometer. Coincident photon detection events in the outputs of the H.O.M. interferometer result in the two spins being projected to an entangled state [3]. Even though QD spin-photon entanglement has now been demonstrated, several important challenges remain to be addressed in order to implement a similar protocol with two QDs.

The spin-photon state associate with decay from the $|T_x-\rangle$ state is a hyper-entangled state which has both spin-frequency and spin-polarization entanglement ($|\Psi\rangle = \frac{|H\rangle|\omega_{red}\rangle|x+\rangle - i|V\rangle|\omega_{blue}\rangle|x-\rangle}{\sqrt{2}}$) [7–9]. First, two QDs are initialized to their $|T_x-\rangle$ state; this state decays, generating two spin-photon entangled states. The QD photons can then be filtered with a quarter-wave plate and polarizer, effectively projecting each $|\Psi\rangle$ to $\sigma_+$, which will result in two spin-frequency entangled states which can be overlapped on the beam splitter of a H.O.M. interferometer [3]. In order for the interference to occur, photons from the two QDs must be indistinguishable [1–3, 10], and even though H.O.M. interference between nearly indistinguishable QD photons has recently been demonstrated [11], efficiently interfacing these measurements with spin manipulation and readout presents several additional challenges.

In practice, it is not uncommon to find two QDs whose trion states have nearly identical bandwidths and lifetimes, and energies that are within tens of GHz of each other. If the QDs are in separate magnetic cryostats, the DC-Stark and Zeeman shifts can provide sufficient tunability to bring these transitions into resonance, even if the g-factors are not matched. Once two "matched" spin-photon entangled states are realized, the remaining challenge is efficient spin readout.

The four-level system of the negatively charged InAs QD in the Voigt geometry provides a long lived spin qubit with fast manipulation through the optically excited trion states [12], but it does not provide a cycling transition for nondestructive spin readout. In the current protocol, the spin readout pulse scatters at most one photon for each shot of the experiment. This pulse reinitializes the system through optical pumping and therefore destroys the information encoded onto the spin state population. In a type-II entanglement protocol, the entanglement generation rate is proportional to the square of the detection efficiency ($DE^2$) since the spin-spin entanglement is heralded on detection of two photons [3, 5], one spin-entangled photon from each QD. In order to verify the spin-spin entanglement between two QDs, two more photons have to be detected to read out the spin states, so the total success rate is proportional to $DE^4$, which is problematic since $DE \ll 1$. Solid immersion lenses and optical cavities have been used to improve light collection from QDs, but are currently limited to $DE \approx 10^{-3}$ for spin-photon interfaces [13]. Given a typical repetition rate of 100 MHz, even though the spin-spin entanglement generation rate is 100 s$^{-1}$, verifying the entanglement with a $DE^4$ protocol will take $\approx 10^2$ hours of integration. One solution is to move to the "W" system of a QD molecule where non-destructive readout of a QD spin state has been demonstrated [14]. This multi-photon readout channel, combined with a triggered readout sequence should provide a near unity readout efficiency, so that the total success rate is limited solely by the entanglement generation rate.

## COMPARISON TO RECENT EXPERIMENTS

After submission of this paper for publication, two related experiments demonstrating QD spin-photon entanglement have been reported. They report entanglement between a single photon, spontaneously emitted from the QD's trion state, and the resulting spin state of the electron confined to the QD [8, 9]. All three of the experiments are consistent in that they report spin-photon entanglement with fidelity lower bounds limited primarily by the timing resolution of the detection system, indicating that near unity fidelities are in principle realizable. Here, we present important physical differences between the experiments.

In the work of De Greve et al., entanglement between the QD electron spin and the polarization state of the photon is reported [8]. Their protocol is similar to this work, but a frequency downconversion technique is used to time gate the single photon detector. The technique is used to improve the timing resolution of the measurement to $< 8$ ps, and to isolate the QD emission from the undesired laser pulses, at the expense of count rate. This enables the use of a larger magnetic field of 3 T, and resulting electron spin precession frequency of 17.6 GHz. An entanglement fidelity bound of $0.8 \pm 0.085$ is reported, which is $84 \pm 10\%$ of the experimentally realizable 0.95, compared to the $84 \pm 6\%$ of experimentally realizable that we report. The $< 8$ ps timing resolution results in a reduced count rate due to the time sampling of the $\approx 600$ ps trion lifetime. Although they report a detection efficiency of $DE \approx 10^{-3}$ without gating, their entanglement generation rate is $2 - 5$ s$^{-1}$ (the current work is $3 \times 10^3$ s$^{-1}$). A primary motivation of these experiments is to realize a scalable quantum network, using the spin-photon entanglement to mediate spin-spin entanglement. Such protocols rely on H.O.M. interference and require indistinguishable photons from each QD, but do not require high (sub-ns) timing resolution. For the commonly used type-II protocols, the spin-spin entanglement generation rate scales with the square of the entanglement generation probability [5]. Given the reported repetition period of a $39 - 52$ ns and the entanglement generation rate of $2 - 5$ s$^{-1}$, the success probability is $< 2 \times 10^{-7}$ [8]. As a consequence, the spin-spin entanglement generation rate using this short pulse downconversion protocol is $< 10^{-6}$ s$^{-1}$. The authors briefly discuss an alternative approach which seeks to overcome this low generation rate, by using a CW downconversion technique to increase the success rate by opening the gate time, while still utilizing the downconversion gating to isolate the QD emission from excitation laser and achieve a telecom wavelength [8].

In Gao et al., entanglement between the QD electron spin and the frequency state of the photon is reported [9], in contrast to spin-polarization entanglement reported here. As in our protocol, polarization is used to block the excitation lasers. First, correlation in the non-rotated basis is demonstrated by measuring the frequency of the QD emitted photon, which is then correlated with the resulting QD spin state. To demonstrate entanglement, it is necessary to show that the correlation is preserved in a rotated basis which is a superposition of the non-rotated basis states. Demonstrating the expected correlation in the rotated basis shows the phase coherence between the qubits. Therefore, in order to verify spin-frequency entanglement, one must measure both the spin qubit and the frequency qubit in a rotated basis. In order to overcome the challenge of performing a rotation of the frequency qubit, the authors use the fact that the polarization state of the emitted photon is correlated with the frequency, based on the Hamiltonian and independent measurements [12]. They perform the rotated basis measurement by passing the entangled photon through a quarter-wave plate and polarizer before detection by a fast single photon detector, which allows for observation of the rotated basis coherence. Instead of projecting the photon onto the two orthogonal rotated basis states ($\sigma_+$ and $\sigma_-$), they rotate the spin by $\pi/2$ and $3\pi/2$ and observe a $\pi$ phase shift in the coincidence oscillation. They report a lower bound on the entanglement fidelity of $0.67 \pm 0.05$, which is $89 \pm 7\%$ of their experimentally realizable 0.75. Their sample design utilizes a solid immersion lens and asymmetric optical cavity to improve light collection. They report a detection efficiency of $\approx 10^{-3}$ [9], which at 76 MHz repetition rate, implies a usable entanglement generation rate of $76 \times 10^3$ s$^{-1}$.

The results of these experiments are in good agreement with result reported here. It is interesting to note that all three experiments yield apparatus corrected fidelities that are nearly identical in terms of the percentage of the maximum achievable. We report that the entanglement fidelity is almost completely limited by the system's timing resolution, which indicates that fidelities exceeding 0.84 should be possible in future studies. It should be noted that sub-ns timing resolution is not required for the two-photon (H.O.M.) interference which will be used for spin-spin entanglement similar to Moehring et al. [3]. While qualitatively verifying the spin-photon entanglement is a necessary step towards quantum information applications of the QD system, ultimately the entanglement fidelity and generation rate between two spins are the figures of merit for future viability of this architecture.

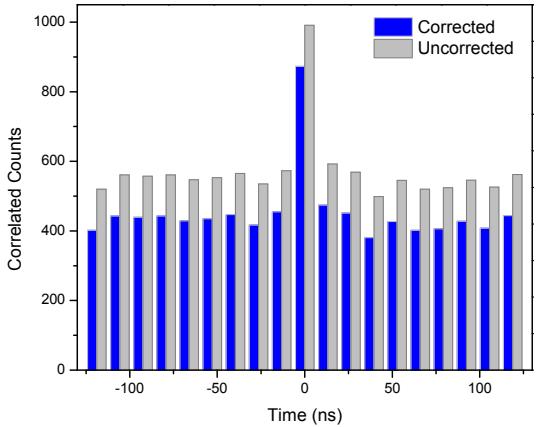

FIG. 1. Correlation data showing positive time zero correlation for the $P(x+|H)$ measurement. The blue (grey) shows the corrected (uncorrected) values over an integration time of 30 minutes.

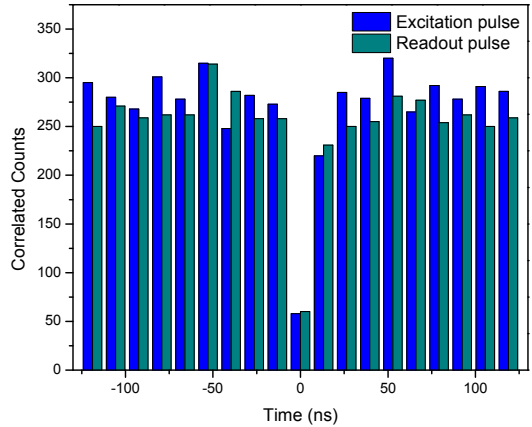

FIG. 2. Correlation data from the $P(x+|H)$ measurement showing the reduced probability of simultaneous detection the two detectors for the same (readout or excitation) pulse. These data are recorded simultaneously with the data shown in Fig. 1. The time-zero value is used to measure the likelihood of emitting two photons from the same pulse of the experiment due to off resonant coupling of the pulse to the other co-polarized transition. The events contribute an offset that is subtracted from each bin of the correlation data (shown in Fig. 1).

## CORRELATION MEASUREMENT IN X BASIS

For each of the conditional probabilities ($P(x-|V)$, $P(x+|V)$, $P(x+|H)$, and $P(x-|H)$), an independent measurement is performed. The details of the pulse sequences are described in the main text. The experiments are performed at an repetition rate of 76 MHz, where each run of the experiment is independent. The data are normalized by comparing the number of correlations between the entangled photon and readout photon from the same run of the experiment and comparing this to the number of correlations between distant runs. The two decay channels (from $|T_x-\rangle$ to $|x+\rangle$ and $|x-\rangle$) are equally likely. The system is excited to the $|T_x-\rangle$ state which then decays by emitting a photon. Without a measurement of the emitted photon state, the resulting spin state is described as an equal mixture of $|x+\rangle$ and $|x-\rangle$, so the probability of scattering a readout photon for a distant run, given an entangled photon at time zero, is half as likely. This is used to normalize these random events to a conditional probability of 0.5. An average over the adjacent $\pm 8$ (excluding $\pm 1$) runs of the experiment is used for the normalization. An example of the uncorrected data is shown in Fig. 1.

The number of detection events consisting of one photon on each of the detectors from the same pulse is also recorded. This can be thought of as a type of anti-bunching arising from optical pumping in the system. Once a photon is detected for either excitation or readout pulses, the QD state is ideally projected to the other spin state, which now only off-resonantly

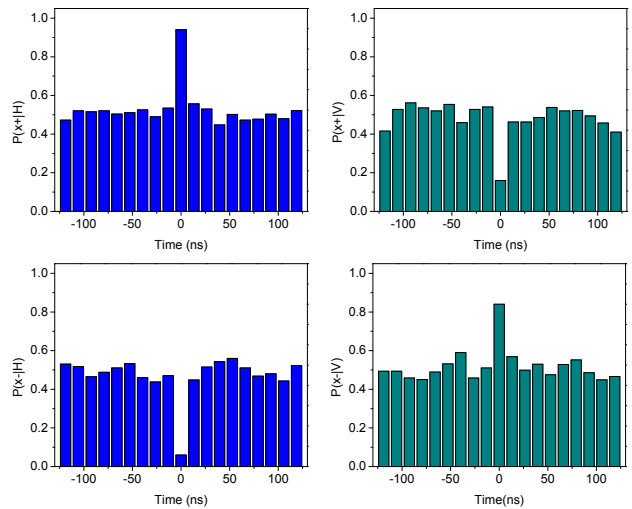

FIG. 3. Normalized correlation events showing spin-photon correlation in the x-basis. The conditional probabilities are: $P(x+|H) = 0.94\pm0.05$, $P(x-|H) = 0.06\pm0.01$, $P(x+|V) = 0.16\pm0.01$, and $P(x-|V) = 0.84\pm0.04$, where the statistical error is $\pm$ one standard deviation.

couples to an excited state. The off-resonant coupling gives a nonzero probability of two photons being emitted from the same pulse and gives rise to a nonzero anti-bunching signal at time zero (Fig 2.). These events provide a measure of the likelihood of emitting a spurious photon from both the excitation and readout pulses, which go like $P(EntangledPhoton) \times P(Error_{ExcitationPulse})$ and $P(ReadoutPhoton) \times P(Error_{ReadoutPulse})$. The values are scaled by the ratio of entangled photons to readout photons to obtain a measure of erroneous correlations between pulses, whose dominant contributions are proportional to $P(EntangledPhoton) \times P(Error_{ReadoutPulse})$ and $P(ReadoutPhoton) \times P(Error_{ExcitationPulse})$. This background contributes equally across time bins so it is subtracted from each run of the raw correlation data prior to normalization. The corrected data are shown in Fig. 1 along with the uncorrected values. While this method partially corrects for the effects of off-resonant coupling leading to spurious correlations, other off-resonant effects reduce the overall measured fidelity (primarily the V polarization measurements) by reducing purity of initialization to the $|T_x-\rangle$ state and cannot be measured directly. The conditional probabilities are then normalized requiring that the sum for each polarization measurement is equal to one. The normalized correlation data for each of the measurements are shown in Fig. 3. Using $\mathcal{F}_x \geq \rho_{Hx+,Hx+} + \rho_{Vx-,Vx-} - 2\sqrt{\rho_{Hx-,Hx-}\rho_{Vx+,Vx+}}$, we obtain a lower bound on the $x$ basis fidelity to be $0.79 \pm 0.03$.

* dst@umich.edu



\end{document}